\documentstyle[twoside]{article}

\catcode`\@=11
\long\def\@makefntext#1{
\protect\noindent \hbox to 3.2pt {\hskip-.9pt
$^{{\eightrm\@thefnmark}}$\hfil}#1\hfill}		

\def\@makefnmark{\hbox to 0pt{$^{\@thefnmark}$\hss}}	

\def\ps@myheadings{\let\@mkboth\@gobbletwo
\def\@oddhead{\hbox{}
\rightmark\hfil\eightrm\thepage}
\def\@oddfoot{}\def\@evenhead{\eightrm\thepage\hfil
\leftmark\hbox{}}\def\@evenfoot{}
\def\sectionmark##1{}\def\subsectionmark##1{}}

\oddsidemargin=\evensidemargin
\newcounter{sectionc}\newcounter{subsectionc}\newcounter{subsubsectionc}
\renewcommand{\section}[1] {\vspace{12pt}\addtocounter{sectionc}{1}
\setcounter{subsectionc}{0}\setcounter{subsubsectionc}{0}\noindent
	{\tenbf\thesectionc. #1}\par\vspace{5pt}}
\renewcommand{\subsection}[1] {\vspace{12pt}\addtocounter{subsectionc}{1}
	\setcounter{subsubsectionc}{0}\noindent
	{\bf\thesectionc.\thesubsectionc. {\kern1pt \bfit #1}}\par\vspace{5pt}}
\renewcommand{\subsubsection}[1] {\vspace{12pt}\addtocounter{subsubsectionc}{1}
	\noindent{\tenrm\thesectionc.\thesubsectionc.\thesubsubsectionc.
	{\kern1pt \tenit #1}}\par\vspace{5pt}}
\newcommand{\nonumsection}[1] {\vspace{12pt}\noindent{\tenbf #1}
	\par\vspace{5pt}}

\newcounter{appendixc}
\newcounter{subappendixc}[appendixc]
\newcounter{subsubappendixc}[subappendixc]
\renewcommand{\thesubappendixc}{\Alph{appendixc}.\arabic{subappendixc}}
\renewcommand{\thesubsubappendixc}
	{\Alph{appendixc}.\arabic{subappendixc}.\arabic{subsubappendixc}}

\renewcommand{\appendix}[1] {\vspace{12pt}
        \refstepcounter{appendixc}
        \setcounter{figure}{0}
        \setcounter{table}{0}
        \setcounter{lemma}{0}
        \setcounter{theorem}{0}
        \setcounter{corollary}{0}
        \setcounter{definition}{0}
        \setcounter{equation}{0}
        \renewcommand{\thefigure}{\Alph{appendixc}.\arabic{figure}}
        \renewcommand{\thetable}{\Alph{appendixc}.\arabic{table}}
        \renewcommand{\theappendixc}{\Alph{appendixc}}
        \renewcommand{\thelemma}{\Alph{appendixc}.\arabic{lemma}}
        \renewcommand{\thetheorem}{\Alph{appendixc}.\arabic{theorem}}
        \renewcommand{\thedefinition}{\Alph{appendixc}.\arabic{definition}}
        \renewcommand{\thecorollary}{\Alph{appendixc}.\arabic{corollary}}
        \renewcommand{\theequation}{\Alph{appendixc}.\arabic{equation}}
        \noindent{\tenbf Appendix \theappendixc #1}\par\vspace{5pt}}
\newcommand{\subappendix}[1] {\vspace{12pt}
        \refstepcounter{subappendixc}
        \noindent{\bf Appendix \thesubappendixc. {\kern1pt \bfit #1}}
	\par\vspace{5pt}}
\newcommand{\subsubappendix}[1] {\vspace{12pt}
        \refstepcounter{subsubappendixc}
        \noindent{\rm Appendix \thesubsubappendixc. {\kern1pt \tenit #1}}
	\par\vspace{5pt}}

\newcommand{\textlineskip}{\baselineskip=13pt}
\newcommand{\smalllineskip}{\baselineskip=10pt}

\def\abstracts#1#2#3{{
	\centering{\begin{minipage}{4.5in}\baselineskip=10pt\footnotesize
	\parindent=0pt #1\par
	\parindent=15pt #2\par
	\parindent=15pt #3
	\end{minipage}}\par}}

\newcommand{\bibit}{\nineit}

\renewenvironment{thebibliography}[1]
	{\frenchspacing
	 \ninerm\baselineskip=11pt
	 \begin{list}{\arabic{enumi}.}
	{\usecounter{enumi}\setlength{\parsep}{0pt}
	 \setlength{\leftmargin 12.7pt}{\rightmargin 0pt} 
	 \setlength{\itemsep}{0pt} \settowidth
	{\labelwidth}{#1.}\sloppy}}{\end{list}}

\newcounter{itemlistc}
\newcounter{romanlistc}
\newcounter{alphlistc}
\newcounter{arabiclistc}

\newcommand{\fcaption}[1]{
        \refstepcounter{figure}
        \setbox\@tempboxa = \hbox{\footnotesize Fig.~\thefigure. #1}
        \ifdim \wd\@tempboxa > 5in
           {\begin{center}
        \parbox{5in}{\footnotesize\smalllineskip Fig.~\thefigure. #1}
            \end{center}}
        \else
             {\begin{center}
             {\footnotesize Fig.~\thefigure. #1}
              \end{center}}
        \fi}

\newcommand{\tcaption}[1]{
        \refstepcounter{table}
        \setbox\@tempboxa = \hbox{\footnotesize Table~\thetable. #1}
        \ifdim \wd\@tempboxa > 5in
           {\begin{center}
        \parbox{5in}{\footnotesize\smalllineskip Table~\thetable. #1}
            \end{center}}
        \else
             {\begin{center}
             {\footnotesize Table~\thetable. #1}
              \end{center}}
        \fi}

\def\@citex[#1]#2{\if@filesw\immediate\write\@auxout
	{\string\citation{#2}}\fi
\def\@citea{}\@cite{\@for\@citeb:=#2\do
	{\@citea\def\@citea{,}\@ifundefined
	{b@\@citeb}{{\bf ?}\@warning
	{Citation `\@citeb' on page \thepage \space undefined}}
	{\csname b@\@citeb\endcsname}}}{#1}}

\newif\if@cghi
\def\cite{\@cghitrue\@ifnextchar [{\@tempswatrue
	\@citex}{\@tempswafalse\@citex[]}}
\def\citelow{\@cghifalse\@ifnextchar [{\@tempswatrue
	\@citex}{\@tempswafalse\@citex[]}}
\def\@cite#1#2{{$\null^{#1}$\if@tempswa\typeout
	{IJCGA warning: optional citation argument
	ignored: `#2'} \fi}}

\def\pmb#1{\setbox0=\hbox{#1}
	\kern-.025em\copy0\kern-\wd0
	\kern.05em\copy0\kern-\wd0
	\kern-.025em\raise.0433em\box0}

\def\fnt#1#2{\footnotetext{\kern-.3em
	{$^{\mbox{\scriptsize #1}}$}{#2}}}

\def\fpage#1{\begingroup
\voffset=.3in
\thispagestyle{empty}\begin{table}[b]\centerline{\footnotesize #1}
	\end{table}\endgroup}

\def\runninghead#1#2{\pagestyle{myheadings}
\markboth{{\protect\footnotesize\it{\quad #1}}\hfill}
{\hfill{\protect\footnotesize\it{#2\quad}}}}
\font\tenrm=cmr10
\font\tenit=cmti10
\font\tenbf=cmbx10
\font\bfit=cmbxti10 at 10pt
\font\ninerm=cmr9
\font\nineit=cmti9

\font\eightrm=cmr8

\def\qed{\hbox{${\vcenter{\vbox{			
   \hrule height 0.4pt\hbox{\vrule width 0.4pt height 6pt
   \kern5pt\vrule width 0.4pt}\hrule height 0.4pt}}}$}}

\def\half{\frac{1}{2}}

\def\metl{g_{ij}}
\def\metlb{\bar{g}_{ij}}
\def\metu{g^{ij}}

\def\eml{F_{ij}}
\def\emu{F^{ij}}

\def\ltf{(\log{\tf})_{uv}}
\def\Lh{\log{h}}

\def\sqd{){\dot{ }}}
\def\apr{a^{\prime}(u)}
\def\bpr{b^{\prime}(v)}
\def\fid{\dot{\phi}}
\def\fids{{\dot{\phi}}^2}
\def\psid{\dot{\psi}}
\def\psids{{\dot{\psi}}^2}
\def\hd{\dot{h}}

\def\fidd{\ddot{\phi}}
\def\be{\begin{equation}}
\def\ee{\end{equation}}
\def\bea{\begin{eqnarray}}
\def\eea{\end{eqnarray}}
\def\tf{\tilde{f}}
\def\th{\tilde{h}}
\def\tV{\tilde{V}}
\def\tVp{{\tilde{V}}^{\prime}}
\def\tYs{\tilde{Y}}
\def\tYf{{\tilde{Y}}_{\phi}}
\def\tYp{{\tilde{Y}}_{\psi}}
\def\dif{\partial}

\def\Hamb{{\bar{\cal H}}^{(s)}}

\def\s{\hspace{1mm}}
\def\EXP{{\rm e}}

\begin{document}

\runninghead{Integrable 1+1 Dimensional Gravity Models}
{Integrable 1+1 Dimensional Gravity Models $\ldots$}

\normalsize\textlineskip
\thispagestyle{empty}
\setcounter{page}{1}


\vspace*{0.88truein}

\fpage{1}
\centerline{\bf INTEGRABLE 1+1 DIMENSIONAL GRAVITY MODELS}
\vspace*{0.37truein}
\centerline{\footnotesize A.T.FILIPPOV}
\vspace*{0.015truein}
\centerline{\footnotesize\it Bogoliubov Laboratory of Theoretical
Physics, Joint Institute for Nuclear Research}
\baselineskip=10pt
\centerline{\footnotesize\it Dubna, Moscow Region, RU-141980, Russia}
\vspace*{0.225truein}

\vspace*{0.21truein}
\abstracts{Integrable models of dilaton gravity coupled to electromagnetic
and scalar matter fields in dimensions 1+1 and 0+1 are reviewed. The 1+1
dimensional integrable models are either solved in terms of explicit
quadratures or reduced to the classically integrable Liouville equation.
The 0+1 dimensional integrable models
emerge as sectors in generally non integrable 1+1 dimensional models
and can be solved in terms of explicit quadratures. The Hamiltonian
formulation and the problem of quantizing are briefly discussed.
Applications to gravity in any space - time dimension are outlined
and a generalization of the so called `no - hair' theorem is proven
using local properties of the Lagrange equations for a rather general
1+1 dimensional dilaton gravity coupled to matter.}{}{}


\vspace*{1pt}\textlineskip	
\section{Introduction}	        
\vspace*{-0.5pt}
\noindent

Many exact solutions of the Einstein - Hilbert (EH) gravity equations were
discovered and studied in the past.
However, it has been only recently realized that these solutions may be
viewed as exact solutions of some integrable Hamiltonian systems  having
much less degrees of freedom than the original EH field theory (for
examples and further references see ref.~\cite{VDA} - ref.~\cite{ATF}).
Typically, 1+1 dimensional field theories or finite systems
(0+1 dimensional field theories)
emerge in this way (similar models are produced by string theories
as well as by topological gauge theories). The Hamiltonian integrability
is very useful in classical theory and it is vital for quantizing.
The idea is that we first exactly quantize the integrable Hamiltonian
subsystem of the full (presumably nonintegrable) system and then apply
a perturbation theory to quantize it. Here we concentrate on the first
step of this program - constructing integrable Hamiltonian subsystems of
the EH gravity coupled to matter.

We will consider here a rather general 1+1 dimensional field theory
which we call dilaton - gravity - matter model (DGM). The gravitational
variables in this model are the metric tensor $g_{ij}$ and the scalar
dilaton field $\phi$. The matter fields are the electromagnetic field
tensor $F_{ij}$ and  the scalar field $\psi$. The DGM Lagrangian is
\be
{\cal L} = \sqrt{-g} \, [U R(g) + V + W\metu \phi_i \phi_j + X\eml \emu +
Y + Z\metu \psi_i \psi_j] \, ,                                   \label{1}
\ee
where $U - Z$ are functions of $\phi$ and $Y$ may in addition depend on
$\psi$; $R$ is the scalar curvature; the lower letter indices denote
partial derivatives ($\phi_i = \partial_i \phi$, etc.), except when used
in $\metl$ or $\eml$. The model may be extended by including many scalar
fields and nonabelian gauge fields; we will not do this exercise here.

With arbitrary functions $U - Z$, this DGM model is certainly not
integrable. However, if $Y = Z = 0$, it is integrable due to local
symmetries discussed below; we will call this important special case
the dilaton gravity (DG) though the Lagrangian includes coupling to
the electromagnetic field. Moreover, the general DG can be exactly reduced
(at least, classically) to a finite - dimensional Hamiltonian system with
one constraint. For the general DGM this is impossible.
The other special case $X = 0$ will be called
the dilaton - gravity - scalar model (DGS). DGS models are generally
not integrable. A new class of integrable 1+1 DGS models as well as DGM
theories having an integrable 0+1 dimensional sectors have been recently
found~\cite{ATF} and will be described below.

The most familiar example of DGM is derived by a dimensional reduction of
 the spherically symmetric sector of the $d$-dimensional
EH gravity coupled to scalar and electromagnetic fields having the same
symmetry. Following ref.~\cite{Soda}, we get the effective Lagrangian in
the form (\ref{1}) with the potentials
\be
U = \EXP^{-2\phi},\;\;V = 2\EXP^{-2\phi}(\alpha \EXP^{4\nu \phi} + \Lambda),
\;\; W = 4(1-\nu)\EXP^{-2\phi},\; \;                             \label{2}
\ee
\be
X = -\EXP^{-2\phi}/\beta ,\;\;\; Y = -\gamma y(\psi)\EXP^{-2\phi},\;\;\;
 Z = -\gamma\EXP^{-2\phi}, \;\;                                  \label{3}
\ee
where $d \equiv n + 2$, $\nu \equiv 1/n$.
Here the parameters $\alpha, \beta, \gamma$ may depend on $\nu$; $\alpha$
is proportional to the curvature of the sphere, $\beta$ and $\gamma$ are
normalization constants. We also have added to the $d$-dimensional action
the cosmological term (by $R \mapsto R + 2\Lambda$). As was pointed
out in ref.~\cite{Soda}, the parameter $\nu$ may formally assume any real
value. In particular, for $\nu = 0$ and $X = Y = =Z = 0$ we obtain
the well - known CGHS dilaton gravity~\cite{CGHS}.
Note that to get the CGHS model coupled to the scalar field we have to
take the constant potential  $Z = -\gamma_0 \neq 0$ and choose
$y(\psi) = 0$.

Many other models (describing black holes, strings, cosmologies)
can be written in the form (1). The approach of our paper may also
be used for coupling of DG to many scalar fields,
nonabelian gauge fields, spinor fields, etc. To compare different models
considered in a rather extensive literature on this subject, one has
to keep in mind that, classically, different parametrizations of
the potentials in terms of $\phi$ and
the Weyl transformations of (1) can be used. For example, for positive
definite $U$ we may use the representation $U \equiv \EXP^{-2\phi}$ or
$U \equiv \phi^2$. For simplicity, we will mainly use $U \equiv \phi$.
In classical theory, it is always possible. In quantum theory,
this parametrization is not necessarily equivalent to the exponential
one or to $U \equiv \phi^2$, etc. (see e.g ref.~\cite{Banks}).
The Weyl transformation $\metl = \Omega (\phi) \metlb$
is even more dangerous. However, in the
classical framework, it can also be used to compare differently looking
models. If two Lagrangians can be identified by using Weyl rescaling
and a different choice for $U$ in terms of $\phi$,
they are classically equivalent.

\vspace*{1pt}\textlineskip
\section{Equations of Motion}
\vspace*{-0.5pt}
\noindent

The equations of motion can be derived by varying ${\cal L}$ with respect
to the variables $\metl$, $\phi$, $\psi$ and $A_i$ (the electromagnetic
potential). Then, using their general covariance, it is convenient to
rewrite them in the conformal flat, light - like metric
$ds^2 = -4f(u,v)du dv$, in which
\be
R = f^{-1}(\log{f})_{uv}.                                      \label{4}
\ee
To further simplify the equations we define
the function $w(\phi)$, $w^{\prime}/w \equiv W/U^{\prime}$
(the prime always denotes the
derivative of a function depending on one variable, thus $U^\prime \equiv
dU/d\phi$, etc.), and write them in the form that is invariant under
the Weyl rescaling, $\metl = \Omega (\phi) \metlb$, i.e.
$\bar{f} = \Omega^{-1} f$. The Lagrangian (1) is invariant if
$\bar{w} = \Omega w$ while $U$, $Z$,
$f w \equiv \tilde{f}$, $V/w \equiv \tilde{V}$, $X w = \tilde{X}$,
$X \eml \emu /w$ and $Y/w \equiv \tilde{Y}$ are  invariant.
Using the fact that the electric charge
$Q = XF_{uv}/2f$ is locally conserved (i.e. $Q_u = Q_v = 0$)
and $X \eml \emu = -2Q^2/X$, we may include the electromagnetic coupling
into the potential $V$ and then forget about the electromagnetic field.
More precisely, we only have to add the term
${\tV}_{em} \equiv 2Q^2 /\tilde{X}$ to the potential $\tV$.

With all these conventions, the equations can be written in the form
\be
\tf (U_i / \tf)_i = Z \psi_i^2\;,\;\;(i = u,v) \, ;               \label{5}
\ee
\be
U_{uv} + \tf (\tV  + \tYs) = 0 \, ;                                \label{6}
\ee
\be
U^{\prime} \ltf +  \tf (\tVp + \tYf) =
Z^{\prime} \psi_u \psi_v\ \, ;                  	           \label{7}
\ee
\be
(Z\psi_u)_v + (Z\psi_v)_u + \tf \tYp = 0 , \;\;
\tYf \equiv \dif_{\phi} \tYs , \;\; \tYp \equiv \dif_{\psi} \tYs . \label{8}
\ee
Not all these equations are independent. Thus, eqs.(\ref{5}) - (\ref{7})
imply (\ref{8}) if $\psi^2_u + \psi^2_v \neq 0$. Similarly, eqs.(\ref{5}),
(\ref{6}), (\ref{8}) imply eq.(\ref{7}) if $\phi^2_u + \phi^2_v \neq 0$.
The statements are easy to prove by considering
$(Z \psi^2_u)_v$ and $(Z \psi^2_v)_u$. A useful corollary is the following.
If $Z^{\prime} = 0$, we only need to solve eqs.(\ref{6}) and (\ref{7}) and
then define $\psi_i^2$ by eqs.(\ref{5}); eq.(\ref{8}) is then automatically
satisfied. These simple observations are very useful for analyzing
integrability. Note that (\ref{7}) is often neglected but it is the
key equation in our approach to integrable DGS models.

\vspace*{1pt}
\section{Integrability of General DG Theory}
\vspace{-0.5pt}
\noindent

If $Y = 0$ and $Z = 0$, we only need to solve the equations (\ref{5})
and (\ref{6}). It is easy to show that they are completely integrable
with arbitrary potential $\tV$. In fact, we can transform them into very
simple linear equations by a B\"acklund - like transformation
\be
M = N(U) + U_u U_v/\tf ,\;\;\Phi_u = \tf / U_v , \;\;
\Phi_v = \tf / U_u \, .                                \label{9}
\ee
The new variables $M$ and $\Phi$ satisfy the linear equations
\be
\Phi_{uv} = 0,\;\;M_u = 0,\;\; M_v = 0.                     \label{10}
\ee
The D'Alembert equation gives the constraint equations (\ref{5}), while
the local conservation of $M$ gives the main `dynamical' equation
(\ref{6}). The solution of the equations (\ref{10}) is obvious,
\be
\Phi = a(u) + b(v) \equiv \tau,\;\;M = {\rm const}.        \label{11}
\ee
To find $U$ and $\tf$ in terms of $M$ and $\tau$ we have first to
express $U$ as a function of $\Phi$ and of $M$, $U = \hat{U}(\Phi , M)$.
It is not difficult to infer from eqs.(\ref{9}) that $\hat{U}$ has to satisfy
the equation $\hat{U}_{\Phi} = M - N(\hat{U})$
which we are free to supplement by the `initial' condition
$\hat{U}(0, M) = 0$. Thus $\Phi =\Phi(\hat{U} , M)$ is defined
by one quadrature and resolving this equation with respect to $\hat{U}$ we
find $U$ as a function of $\tau$ and $M$. Then we have
\be
\tf = \hat{U}_{\Phi} \Phi_u \Phi_v =
[M - N(\hat{U}(\Phi , M))] \Phi_u \Phi_v\,,                   \label{11a}
\ee
which completes the solution ($\th \equiv hw$),
\be
\tf = [M - N(\hat{U}(\tau , M))] \apr \bpr
\equiv \th(\tau , M) \apr \bpr \,.    				\label{12}
\ee
Thus, $\phi$ depends on one coordinate $\tau$. Choosing $a$ and $b$ as
the new coordinates, we find that the metric also depends on one coordinate,
\be
ds^2 = -4 f du dv = -4 h da db.                             \label{12a}
\ee
We see that the solution of the equations of motion essentially depend on
one variable $\tau = a + b$.
It is called `static' because it is a generalization of
the static Schwarzschild solution in general relativity (black hole).
Then zeroes of $h(\phi) \equiv h(U(\phi))$
for finite values of $\phi$ are called the horizons; with one horizon, we
have a Schwarzschild - like black hole solution and $M$ is the mass of
the black hole.

We will not further elaborate the physics interpretation
of the solution and will not give the Hamiltonian formulation of DG
in dimension 1+1. We may simply use the fact that the solutions of the
DG equations are essentially 0+1 dimensional and write the corresponding
constrained Hamiltonian equations. This will be shown for the static
sectors of the DGM models which, in general, are not integrable and can
not be completely reduced to 0+1 dimension.

\vspace*{1pt}\textlineskip	
\section{Integrable DGS Models}	        
\vspace*{-0.5pt}
\noindent

The 1+1 dimensional integrable models discussed below classically
reduce to the Liouville equation (to understand why, please look at
the expression for the two dimensional scalar curvature (\ref{4})):
\be
G_{uv} + 2g \exp{G} = 0.					\label{13}
\ee
Its general solution may be found by simple B\"acklund transformations
of the Liouville equation to the D'Alembert equation $\Phi_{uv} = 0$,
for example,
\be
\exp{G} = - g^{-1} \Phi_u \Phi_v / \Phi^2 \, .           \label{14}
\ee
This and other representations of $G$ in terms of the solutions of the
D'Alembert equation may be obtained from the following simple observation
(known to Liouville!): if $G^0 (u,v)$ is a solution of (\ref{13}) then
\be
G^1 (u,v) = G^0 (a(u),b(v)) + \log(\apr \bpr)           \label{15}
\ee
satisfy the same equation for arbitrary functions $a(u)$ and $b(v)$.
Now, taking $G^0$ defined by $\exp{G^0} = \pm g^{-1} (u \mp v)^{-2}$,
where $\pm g > 0$, we find that $G^1$ can be represented in the form
(\ref{14}). Substituting $\tan a$ and $\tan b$ for $a$ and $b$, we
may find another representation for $G$,
\be
\exp{G} = - g^{-1} \Phi_u \Phi_v / \sin^2{\Phi} \, .       \label{16}
\ee
The important point is that the solutions of the Liouville equation
essentially depend on one variable $\Phi = a(u) + b(v)$ and that, in
general, they have singularities in this variable. This poses serious
problems in their geometric interpretation as well as in quantizing.
To the best of my knowledge, these problems are not yet solved.
Note that the class of the solutions without singularities
may be represented in the form
\be
\exp{G} = g^{-1} \Phi_u \Phi_v / \cosh^2{\Phi} \, ,       \label{17}
\ee
which can be derived by substituting $\tanh{a}$ and $\coth{b}$ for
$a$ and $b$. One may try to approach the mentioned problems by
studying first this non singular sector of the Liouville equation.

Now, let us write integrable DGS models with $Z^{\prime} = 0$ (due to
their relation to string theory they are often called string motivated
dilaton gravity models). Without restricting the generality of our
consideration, we will choose the parametrization $U = \phi$ (then
$U^{\prime} \equiv 1$). Adding eq.(\ref{6}) multiplied
by constants $\pm g_1$ to eq.(\ref{7}), we find
\be
F_{uv}^{\pm} + \epsilon \, [(\tilde{V}^{\prime}\pm g_1 \tilde{V})
\EXP^{\mp g_1 \phi}] \exp{F^{\pm}} = 0\, ,              \label{18}
\ee
where $F^{\pm} = \log{(\tilde{f} \EXP^{\pm g_1 \phi})}$ and $\epsilon$
is the sign of
$\tilde{f}$. Now, if we choose $\tilde{V}$ so that the expression in the
square brackets is constant, $2 g_{\pm}$, these equations
coincide with the Liouville equation.
The most general potential satisfying this requirement is
\be
\tilde{V} = (g_{+} \EXP^{g_1 \phi} -
g_{-} \EXP^{-g_1 \phi})/g_1 \, .                               \label{19}
\ee
Using the above formulas we may reduce
eqs.(\ref{18}) to two D'Alembert equations
${\Phi}^{\pm}_{uv} = 0$ and thus solve
them in terms of four arbitrary functions $a^{\pm}(u)$, $b^{\pm}(v)$.

A simpler integrable model is defined by the potential
\be
\tV = g_3 + 2g_2 \phi \, \;\; g_2 \equiv g_{+} \, ,             \label{20}
\ee
which can be obtained from (\ref{19}) with
$g_{-} \equiv g_{+} - g_3 g_1$ and $g_1 \to 0$.
Then (\ref{7}) is the Liouville equation for the metric while (\ref{6})
is the linear equation for $\phi$ defining scattering of the dilaton
field on the metric. Both equations can be explicitly solved. Note that
there is no scattering on nonsingular metrics defined by the solutions
of the Liouville equation (\ref{17}). To see this, define the new
coordinates $r = a + b$ and $t = a - b$. Then the equation for the
dilaton field is the Klein - Gordon equation with the reflectionless
potential $2/\cosh^2(r)$. A detailed discussion of this and other
solutions will be presented elsewhere.

The obtained integrable DGS models significantly generalize the
CGHS model, in which the curvature $R$ is zero. In our model
(\ref{20}), it is constant while in the most general integrable model
(\ref{19}) it is not constant. Even on the classical level, constructing
global solutions of these models is an unsolved problem. I think that
the singularities signal a necessity of considering topologically
nontrivial space - time surfaces. In quantum theory, some formal operator
solutions of the Liouville equation are known~\cite{Andrei}. Whether
these solutions allow a Hilbert space realization and can give
a complete quantum theory of the integrable DGS models is at the moment
unclear.

\vspace*{1pt}\textlineskip	
\section{Integrable 0+1 Dimensional Sectors of DGM Models}
\vspace*{-0.5pt}
\noindent

All DGM model obtained by the dimensional reduction of the EH gravity
coupled to scalar fields have $Z^{\prime} \neq 0$.
However, I do not know any
integrable DGS model with $Z^{\prime} \neq 0$ in dimension 1+1.
Nevertheless, for the spherically symmetric sector of the EH gravity
coupled to scalar fields several exact solutions are known for long time.
In our language, these solutions generate integrable 0+1 dimensional
constrained systems which form certain sectors in 1+1 dimensional DGM.
We will call these sectors `static'. Let us define what is the static
solution of DGM not using these intuitive argument and in more precise
terms.  We call the solution of eqs.(\ref{5}) - (\ref{8})
{\it static} if there exist functions $a(u)$ and $b(v)$ such that
$\phi$, $h \equiv f/\apr \bpr$ and $\psi$ depend only on
$\tau \equiv a + b$
(in view of the above considerations, this definition is quite natural).
In fact, if $\phi = \phi(\tau)$ and $\psi_a + \psi_b \neq 0$,
the equations (\ref{6}), (\ref{5}) tell us
that $\psi = \psi(\tau)$, $h = h(\tau)$ and thus the solution is static.
The set of all static solutions forms the static sector of a given DGS
model.

The static sector is a 0+1 dimensional theory and we will show that
$\phi(\tau)$, $\psi(\tau)$ and $h(\tau)$
are coordinates of a constrained Hamiltonian system (the constraint is the
Hamiltonian $H$ itself, i.e. $H = 0$). This system is integrable if there
exist two more integrals depending on the coordinates $h$, $\phi$, $\psi$
and velocities $\hd$, $\fid$, $\psid$ (or momenta introduced below)
and the system of the first-order
equations defined by these integrals is explicitly integrable (recall that
the omitted equation for the electromagnetic field is always integrable).
Then we call the DGM model {\it s-integrable}. If $Y = 0$,
the equation (\ref{8}), that now is simply $(Z \psid \sqd = 0$,
gives the integral $C_0 = Z \psid$. Thus the problem is to find one more
integral. Keeping this in mind, let us rewrite the remaining
equations (\ref{5}) - (\ref{7}) for the static sector.
Then we have the following ordinary differential equations
(we now take $U \equiv \phi$ and return to the Weyl - nonivariant notation):
\be
h w (\fid /h w\sqd \equiv
\fidd - W \fids - \fid F = Z \psids \, ;                    \label{21}
\ee
\be
\fidd + h V = 0\, ;                                         \label{22}
\ee
\be
2W \fidd + W^{\prime} \fids + \dot{F} +
h V^{\prime} = Z^{\prime} \psids  \,                        \label{23}
\ee
(for convenience, we introduce a temporal notation $F \equiv \dot{h} / h$).

The Hamiltonian constraint can be directly obtained from (\ref{21})
and (\ref{22})
\be
L \equiv W \fids + \fid F  + h V + Z \psids = 0,
\;\; F \equiv \dot{h} / h \;.                                    \label{24}
\ee
As was shown in ref.~\cite{ATF}, an additional integral can be found
in the following two cases.
If $Z$ and $\tV$ satisfy the relation
(recall that $N^{\prime}(\phi) \equiv \tV$):
\be
Z = (g_2 + g_1 N(\phi)) {\tV}^{-1}  ,                            \label{25}
\ee
where $g_1$, $g_2$ are real constants, we have the integral
\be
Z F + (Z W - g_1) \fid = C_1\s .                                  \label{26}
\ee
The three available integrals $C_0$, $C_1$ and $L$ allow us to
find the general solution to all the equations in terms of explicit
quadratures.

The second s-integrable case is given by two relations for the potentials
\be
V = W({\bar{g}}_4 w^2 - {\bar{g}}_1),\;\;\; Z^{-1} = W ({\bar{g}}_3 +
{\bar{g}_2} \log{w}),                                             \label{27}
\ee
where ${\bar{g}_1}$ - ${\bar{g}_4}$ are arbitrary real constants.
The additional integral of motion is
\be
(F/W)^2 + 4 {\bar{g}}_1 h + 2 {\bar{g}}_2 C_0^2 \Lh = {\bar{C}}_1\ .
                                                                  \label{28}
\ee
As in the first case, this integral allows to write the solution of all
the equation in terms of explicit quadratures.

\vspace*{1pt}\textlineskip
\section{Hamiltonian Formulation of 0+1 Dimensional Models}
\vspace{-0.5pt}
\noindent

Let us now give the Hamiltonian formulation of
the `static' equations of DGM. It is not difficult to show that
Eqs.(\ref{21}) - (\ref{23}) as well as the constraint (\ref{24}) and
the omitted equation for $\psi$ can be derived by varying the Lagrangian
\be
{\cal L}^{(s)} \equiv (L - hV)/l - lhV                         \label{29}
\ee
in all variables including the Lagrangian multiplier $l(\tau)$.
Then the constraint equation (\ref{24}) is reproduced for $l = 1$,
which is nothing but a choice of a gauge.
The necessity of this gauge fixing is
related to the fact that we are using the conformal flat metric
depending on one function $f(u,v)$.
If we would use a more general metric, e.g.
 $ds^2 = \alpha(r) dr^2 - \beta(r) dt^2$,
a Lagrangian multiplier will emerge automatically.

Introducing the momenta
\be
p_{h} = \fid/lh \, , \;\;p_{\phi} = (2W \fid + \hd /h)/l \, , \;\;p_{\psi} =
2Z \psid /l \, ,                                              \label{30}
\ee
we find the Hamiltonian ${\cal H}^{(s)} = l H$, where
\be
H = h p_h p_{\phi} - W h^2 p_h^2 + h V + p_{\psi}^2/4 Z \, .   \label{31}
\ee
Now one may express the integrals of motion in terms of the canonical
coordinates and momenta and check that their Poisson brackets with
$H$ vanish when these variables satisfy the canonical equations
of motion (including the constraint $H = 0$). Of course, the
canonical equations are equivalent to the Lagrangian ones that were solved
above (with $l = 1$).

Note that the form of the Hamiltonian (and of the Lagrangian) is not
uniquely defined, due to a freedom in the choice of the Lagrangian
multiplier $l(\tau)$. We may multiply $H$ by a function
of the coordinates, $\lambda(h, \phi, \psi)$ (not having zeroes inside
the domain of definition of the coordinates) and correspondingly divide
$l(\tau)$ by $\lambda$. The new  Hamiltonian
$\Hamb \equiv \bar{l} \bar{H}$ define the same equations of
motion due to the constraint  $H = 0$. Such a freedom
is useful because the new Hamiltonian may have additional integrals of
motion. This observation was used in our approach to quantizing
black holes~\cite{VDA}. In that case, the mass of the black hole
is proportional to $M$ defined in (\ref{9}). It is conserved
when the scalar field is completely decoupled, i.e. for $C_0 = 0$.
For the `static' case, the mass function is simply
\be
M = N(\phi) + h p_h^2 /w \equiv N(\phi) + \fids /h w \ .  \label{32}
\ee
When $p_\psi \equiv 2 C_0 \neq 0$, $M$ is not conserved but, for s-integrable
models, there may exist other integrals of motion~\cite{Marco}. In the
s-integrable models, the integrals $C_1$ and ${\bar{C}}_1$ found in this
paper play the role of $M$. Thus, it is not difficult to show that
\be
C_1 = -g_1 {w \over p_h} \left( M + {g_2 \over g_1} +
{p_\psi^2 \over 4 g_1 h w} \right).                        \label{33}
\ee
When $p_\psi \equiv 2 C_0 = 0$, the factor $w/p_f$ in (34) becomes an
additional integral of motion because $C_2 \equiv p_h /w = \fid /h w$ is
independent of $\tau$ due to (\ref{21}). A relation similar to (\ref{33})
may be derived for ${\bar{C}}_1$. We write it only for $p_\psi = 0$:
\be
{\bar{C}}_1 = (M^2 - 4 {\bar{g}}_1 {\bar{g}}_4)/ C_2^2 \ . \label{34}
\ee

Note that the above Lagrangian and Hamiltonian formulations are valid
in general, when additional integrals of motion are not known or
even do not exist. Unfortunately, for nonlinear Hamiltonian systems
with two or more
independent coordinates the existence of such an integral is a rare
event. Apparently simple systems with two coordinates are not
integrable and thus exhibit complex phenomena known as dynamical
chaos. A famous example is the Henon - Heiles system of two oscillators
with cubic couplings between them. If one
compares our general `static' system (\ref{31}) (with fixed $p_\psi$
and arbitrary functions $V$, $W$ and $Z$)
to such well-studied nonitegrable systems, one may infer that it must
be not integrable (one may show that, due to the Weyl invariance and
the constraint $H = 0$,
the general model essentially depends only on one arbitrary function but
this does not help).

In particular, this means that the general DGM theory is not integrable.
This does not mean that some other examples of integrable DGM models
can not be discovered in future.
On the other hand, chaotic phenomena in classical
nonintegrable DGM models might be of significant physics interest.
In quantum framework, nonintegrable models may still be useful if
they can be treated perturbatively on some explicitly integrable
background (like our s-integrable models).

\vspace*{1pt}\textlineskip
\section{An Example and a Theorem}
\vspace*{-0.5pt}
\noindent

Here we give a very simple static solution for DGM that shows why the
scalar field can not be treated as a perturbation. We also will argue
that the static metric for general DGM (integrable or not) have no
horizons. This is a new version of the so called
`no-hair' theorem~\cite{Beken}.

The model described by the potentials (\ref{27}) with ${\bar{g}}_1 =
{\bar{g}}_2 = 0$ is very easy to integrate. By solving the equations
of motion we find
\be
h_0 w = ({\bar{g}}_4 / {\bar{C}}_1)^{-{\half}} \, (1 + 2\delta) \,
|h/h_0|^{\delta} \, \bigl| 1 +
\epsilon |h/h_0|^{1+2\delta}\bigr|^{-1} \ ,                  \label{35}
\ee
where $h_0$ is the integration constant, $\epsilon$ is the sign of $h$ and
\be
2\delta = \sqrt{1 - 4 {\bar{g}_3} C_0^2 / {\bar{C}}_1} - 1 \ .  \label{36}
\ee
If $h \to 0$, then $w \to 0$ or $w \to \infty$. For the dimensionally
reduced gravity defined by (\ref{2}) (with $\Lambda =0$)
the coupling constant ${\bar{g}_3}$ is negative and thus $w \to 0$,
$\phi \to 0$. The metric has no zeroes for finite values of $\phi$, i.e.
no horizons. When $C_0^2 = 0$ and thus $\delta = 0$, the scalar
field disappears and the horizon reappears. However, there is no analytic
transition $\delta \to 0$. This means that the scalar field can not
be treated as a perturbation.

Now we will show that the absence of zeroes in $h$ for finite values
of $\phi$ is a general property of DGM. To prove this, rewrite
(\ref{21}) and (\ref{22}) as follows
\be
\dot{\phi} = \chi , \;\; \dot{\chi} = -hV, \;\; \dot{h} =
-h {\chi}^{-1} (C_0^2 Z^{-1} + hV).                              \label{37}
\ee
As the right-hand sides of these equations are independent of $\tau$,
we can reduce these three equations to two equations for $\chi$ and
$\phi$ as functions of $h$. Let us consider the behavior of
$\chi$ and $\phi$ for $h \to 0$ assuming that $\phi \to {\phi}_0
\neq \infty$. Suppose also that, in this limit, $V \to V_0 \neq \infty$
and $Z \to Z_0 \neq 0, \ \infty$. Then it is easy to find that
\be
{d \log{\chi} \over dh} \to {Z_0 V_0 \over C_0^2} \ ,              \label{38}
\ee
which means that $\chi \to {\chi}_0 \neq 0, \ \infty$. Now one can
show that
\be
{d\phi \over d\log{h}} \to -{{\chi}_0^2 Z_0 \over C_0^2} \ .      \label{39}
\ee
It follows that $|\phi| \to \infty$. This contradicts our assumption
that $\phi_0 \neq \infty$.

In the picturesque language used in the gravity literature,
this means that static black hole solutions of the Einstein
gravity have no scalar `hair'. The precise formulation of the theorem
we have just proven is: if $Z \neq 0$, the static solutions of the
general DGM theory have no horizons (in our proof we supposed also
that $Y = 0$). Further generalizations and refinements of this theorem
will be published elsewhere. Note that, in this formulation, the
`no-hair' theorem is a local property of the differential equations
of motion.


\vspace*{1pt}\textlineskip
\section{Summary}
\vspace{-0.5pt}
\noindent

We have studied a general dilaton gravity coupled to electromagnetic
and scalar fields, \ref{1}. For the general $U - X$ model (DG) we presented
some known results in a more compact and simple form emphasizing its
integrability. For the $U - Z$ models (DGS) we concentrated on a somewhat
simpler case $Y = 0$ and obtained a class of integrable theories
with the constant potential $Z$, thus generalizing the CGHS dilaton
gravity. Note that there exist integrable models with $Y \neq 0$ which
we will present elsewhere. For the DGM theories with arbitrary $Z$, we
found two classes of s-integrable (and explicitly soluble) systems.
A special case of the s-integrable DGM gives new exact solutions of the
Einstein gravity coupled to matter in any space - time dimension.
We also pointed out a generalization of the `no - hair' theorem.
By constructing the Hamiltonian formulation of the s-integrable systems
we hopefully paved a way to their quantizing. Further applications to
gravity will be presented elsewhere.

\nonumsection{Acknowledgments}
\noindent
Useful discussion of the ideas presented above with L.D.Faddeev,
A.N.Leznov and V.A.Rubakov are kindly acknowledged.
This investigation was partially supported
by the Russian Fundamental Science Foundation (project 95-02-05679),
and by INTAS (project 93-0127).

\nonumsection{References}

\end{document}